\documentclass[aps,twocolumn,groupedaddress,showpacs,amsmath]{revtex4}

\usepackage{epsf,latexsym}
\usepackage{amssymb,amsmath}
\usepackage{times}

\newcommand{\ket}[1]{| #1 \rangle}

\newcommand{\ignore}[1]{}

\newcommand{\ra}{{\rightarrow}}
\newcommand{\up}{{\uparrow}}
\newcommand{\down}{{\downarrow}}

\newcommand{\be}{\begin{equation}}
\newcommand{\ee}{\end{equation}}
\newcommand{\ba}{\begin{eqnarray}}
\newcommand{\ea}{\end{eqnarray}}

\def\CC{{\rm\kern.24em \vrule width.04em height1.46ex depth-.07ex
    \kern-.30em C}}
\def\P{{\rm I\kern-.25em P}}

\def\RR{{\rm
         \vrule width.04em height1.58ex depth-.0ex
         \kern-.04em R}}

\def\bbbone{{\mathchoice {\rm 1\mskip-4mu l} {\rm 1\mskip-4mu l}
{\rm 1\mskip-4.5mu l} {\rm 1\mskip-5mu l}}}

\def\bbbc{{\mathchoice {\setbox0=\hbox{$\displaystyle\rm C$}\hbox{\hbox
to0pt{\kern0.4\wd0\vrule height0.9\ht0\hss}\box0}}
{\setbox0=\hbox{$\textstyle\rm C$}\hbox{\hbox
to0pt{\kern0.4\wd0\vrule height0.9\ht0\hss}\box0}}
{\setbox0=\hbox{$\scriptstyle\rm C$}\hbox{\hbox
to0pt{\kern0.4\wd0\vrule height0.9\ht0\hss}\box0}}
{\setbox0=\hbox{$\scriptscriptstyle\rm C$}\hbox{\hbox
to0pt{\kern0.4\wd0\vrule height0.9\ht0\hss}\box0}}}}

\def\bbbz{{\mathchoice {\hbox{$\sf\textstyle Z\kern-0.4em Z$}}
{\hbox{$\sf\textstyle Z\kern-0.4em Z$}}
{\hbox{$\sf\scriptstyle Z\kern-0.3em Z$}}
{\hbox{$\sf\scriptscriptstyle Z\kern-0.2em Z$}}}}

\newcommand{\putfig}[2]{$$\leavevmode\hbox{\epsfxsize=#2 cm
   \epsffile{#1.eps}}$$}
\newcommand{\insertfig}[2]{\leavevmode \vcenter{\hbox{\epsfxsize=#2 cm
   \epsffile{#1.eps}}}}

\begin{document}

\title{Entangling spins by measuring charge: a parity-gate toolbox}

\author{Radu Ionicioiu}
\affiliation{Quantum Information Group, Institute for Scientific Interchange (ISI), Viale Settimio Severo 65, I-10133 Torino, Italy}
\altaffiliation{Present address: Hewlett-Packard Laboratories, Filton Road, Stoke Gifford, Bristol BS34 8QZ, United Kingdom}

\begin{abstract}
The parity gate emerged recently as a promising resource for performing universal quantum computation with fermions using only linear interactions. Here we analyse the parity gate ($P$-gate) from a theoretical point of view in the context of quantum networks. We present several schemes for entanglement generation with $P$-gates and show that native networks simplify considerably the resources required for producing multi-qubit entanglement, like $n$-GHZ states. Other applications include a Bell-state analyser and teleportation. We also show that cluster state fusion can be performed deterministically with parity measurements. We then extend this analysis to hybrid quantum networks containing spin and mode qubits. Starting from an easy-to-prepare resource (spin-mode entanglement of single electrons) we show how to produce a spin $n$-GHZ state with linear elements (beam-splitters and local spin-flips) and charge-parity detectors; this state can be used as a resource in a spin quantum computer or as a precursor for constructing cluster states. Finally, we construct a novel spin $CZ$-gate by using the mode degrees of freedom as ancill\ae.
\end{abstract}


\maketitle

\section{Introduction}

One of the breakthrough insights in quantum information has been the understanding that measurement can provide the nonlinearity required for implementing two-qubit gates \cite{klm}. This result runs contrary to the common intuition that a unitary (hence reversible) gate cannot be constructed out of irreversible (hence non-unitary) operations. The pioneering work of Knill, Laflamme and Milburn (KLM) \cite{klm} changed completely the prevailing paradigm and paved the way to measurement-based approaches of quantum computation (QC) \cite{nielsen,leung}. A different measurement-based idea is the one-way quantum computation model introduced by Raussendorf and Briegel \cite{1wqc}, in which the computation is performed by single-qubit measurements on a highly entangled initial state (the cluster state).

The key resource in the KLM model are photon number discriminating detectors which can distinguish between zero, one and two photons. This imposes severe experimental restrictions and are still difficult to implement in practice. Conceptually, the CNOT gate in the KLM model uses the bosonic nature of photons, i.e., bunching at a beam-splitter. The impact of the KLM model motivated the search for a similar construction for fermions, but the initial efforts have been hampered by no-go theorems \cite{terhal, knill}. The results obtained by Terhal, DiVincenzo and Knill show that for fermions single-electron Hamiltonians and single-spin measurements can be efficiently simulated classically. This obviously implies that the exponential speed-up giving the edge of quantum over classical computation cannot be realized, in the fermionic case, with only linear interactions and single-spin measurements.

The story, however, does not end here. Another landmark result has recently reopened the debate of fermionic QC with linear elements. Beenakker {\it et al.} \cite{spin_parity} showed how to sidestep the previous no-go theorems by using charge parity measurements instead of single-spin measurements. The key element in Ref.~\cite{spin_parity} is the construction of a CNOT gate for spins using parity measurements, an ancilla and post-processing. Soon thereafter implementations for parity gate with spin \cite{engel,kolli,coish} and charge qubits \cite{charge_parity} have been proposed in the literature.

Although the initial focus was on solid state implementations (mainly spins in quantum dots), the importance of the parity gate ($P$-gate) extends beyond this framework. In this article we take a different route, by investigating the $P$-gate from an abstract point of view, without resorting to a particular implementation. Our main tool will be quantum network analysis, extending the results in \cite{ri}. First, we analyse different quantum networks containing $P$-gates and we derive gate identities. We then study entanglement generation with parity measurements and present a non-destructive Bell state analyser; this can be used as a primitive for a teleportation network. In the same time we show how ``native'' quantum networks designed for $P$-gates can simplify considerably the resources needed for generating multi-qubit entangled states ($n$-GHZ states), compared to the traditional approach of building first CNOT gates out of $P$-gates. Another important application of parity measurements is in performing a deterministic cluster state fusion.

We then extended this approach to hybrid quantum networks containing both spin and mode qubits. In this case we show a novel method of entangling spins starting from an easy-to-prepare resource, i.e., single-particle entanglement between spin-mode degrees of freedom. With this method we can produce a spin $n$-GHZ state with linear elements (beam-splitters and local spin-flips), plus charge-parity measurements. In the same way one can prepare {\em hyper-entangled} states (entangled in both spin and mode) of $n$ electrons. Another outcome of this approach is a novel spin $CZ$-gate which uses the mode degrees of freedom as ancill\ae.

\section{The parity gate}

\subsection{Notations and overview}

In the following we denote by $X_i$, $Z_i$ the usual Pauli operators $\sigma_{x,z}$ acting on the $i$-th qubit. For a binary vector ${\bf v}= (v_1,..,v_n)\in \bbbz_2^n$, we define $X({\bf v}):= \prod_{i=1}^n X_i^{v_i}$ and analogously for $Z({\bf v})$. Let $X^{\otimes n}:= \prod_{i=1}^n X_i$ be the operator flipping the state of all $n$ qubits. Given a basis vector of $n$ qubits $\ket{j}$, with $j=0,..,2^n-1$, we can regard it as a binary vector ${\bf j}=(j_1,..,j_n) \in \bbbz_2^n$ and we have $\ket{j}= X({\bf j}) \ket{0}^{\otimes n}$. 

\begin{figure}
\putfig{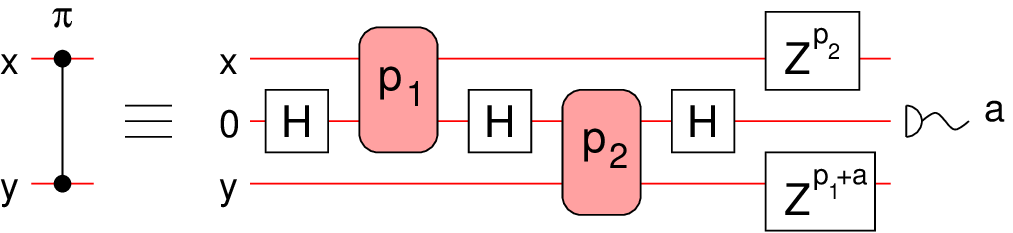}{8}
\putfig{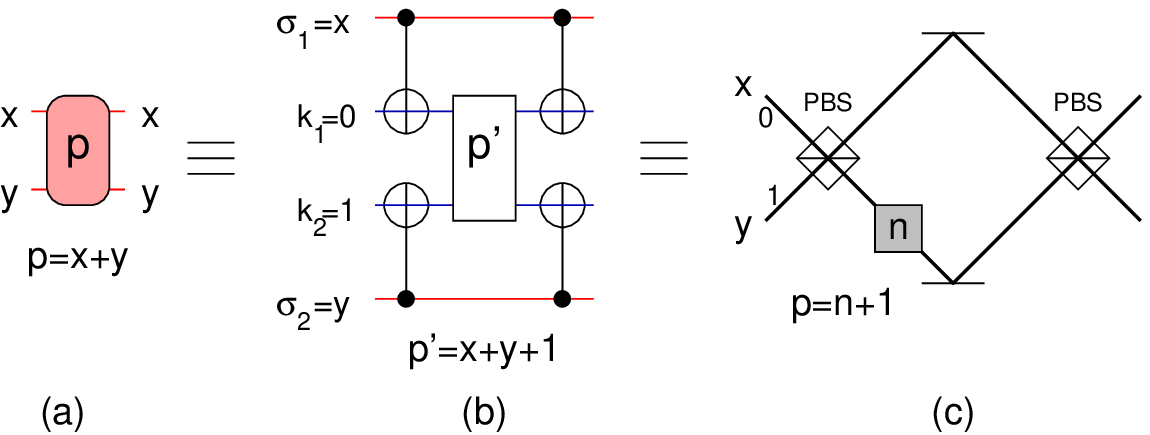}{8}
\caption{(color online) Top: A quantum network for a spin $CZ$ gate using two spin-parity measurements (pink boxes). Bottom: each spin-parity gate (a) is constructed out of a charge-parity gate (white box) via charge-to-spin conversion (the CNOT gates between spin and charge qubits) as in (b). Implementation of the previous circuit using ballistic electrons (c), see Ref.~\cite{spin_parity}; the grey box is a non-absorbing charge parity detector measuring the parity $n$ in the lower arm of the interferometer; all sums are mod 2.}
\label{cz_gate}
\end{figure}

In order to make the connection with previous work, we first review the construction of Ref.~\cite{spin_parity}. The main resource is a charge-parity detector measuring the particle number modulo 2 in a given spatial mode, $p:= Q \mod 2$; here $Q:= n_\up+ n_\down$ is the total charge operator. With this as a primitive, one first construct a spin-parity gate via charge-to-spin conversion (using two polarizing beam-splitters), as in Fig.~\ref{cz_gate}, bottom. The next step is to use the spin-parity detector in order to build a spin-CNOT gate.

In Fig.~\ref{cz_gate}, top, we show the equivalent construction of a $CZ$ (controlled-$Z$) gate. The only differences with respect to the original construction \cite{spin_parity} are two Hadamard gates on the target qubit and a redefinition of the parity measurement output: in our case the $P$-gate measures the parity $p= x\oplus y$ of the input qubits instead of the particle number $n$ after PBS; the two are equivalent, since $p=n\oplus 1$ ($\oplus$ is the addition mod 2).

There are two important points of this construction. First, the gate is deterministic (in contrast to the KLM model, which implements a probabilistic $CZ$ gate). There is a post-processing stage -- the final $Z^p$ gates on the computational qubits -- ensuring the correct output irrespective of the measured values $p_1,p_2,a$. Second, the gate works {\em coherently} on superposition states, hence it can be used in general quantum networks \cite{spin_parity}.

\subsection{Quantum networks and gate identities}

In order to have a better understanding of the parity gate, in this section we analyse quantum networks containing the $P$-gate and we derive gate identities.

First, let us have a look at how the $P$-gate acts on a general two-qubit state $\ket{\psi}= a\ket{00}+ b\ket{01}+ c\ket{10}+ d\ket{11}$. Upon a parity measurement, $\ket{\psi}$ is projected to one of the two subspaces of equal parity, i.e., the (unnormalized) output state is either $a\ket{00}+ d\ket{11}$ (with probability $|a|^2+ |d|^2$) for parity $p=0$, or $b\ket{01}+ c\ket{10}$ (with probability $|b|^2+ |c|^2$) for $p=1$. Clearly, the $P$-gate leaves invariant the basis states $\ket{xy}$, $x,y=0,1$ and outputs only the parity $p=x\oplus y$.

\begin{figure}
\putfig{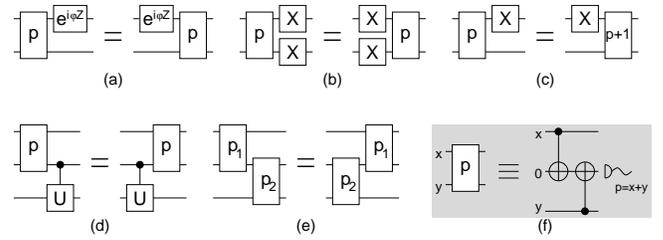}{8.5}
\caption{Gate identities for the parity gate. The $P$-gate commutes with: (a) general $Z$-rotations $e^{i\varphi Z}$, (b) global spin-flips $X_1\otimes X_2$, and (d) arbitrary controlled-$U$ gates. (c) Commuting with $X$ changes the value $p$ of the parity into $p\oplus 1$. Since the $P$-gate is symmetric in the two inputs, all the commutation relations work on both qubit lines. (e) Two $P$-gates commute when acting on three qubits. (f) A quantum network model for the $P$-gate; the ancilla is symmetrically coupled to the two qubits and then measured, giving the parity of the input states $p=x+y\mod 2$.}
\label{gate_id}
\end{figure}

A useful tool in constructing general quantum networks are gate identities, i.e., relationship between equivalent networks, like commutation relations between various gates. Gate identities are indispensable for transforming and simplifying quantum networks and for gaining insight of their functionality. Using the previous action, it is easy to derive gate identities for the $P$-gate and these are given in Fig.~\ref{gate_id}. The $P$-gate commutes with general $Z$-rotations $e^{i\varphi Z}$ and arbitrary controlled-$U$ gates. Commuting with $X$ gate flips the parity, $p\mapsto p\oplus 1$. Since the $P$-gate is completely symmetric in the two inputs, the same gate identities work on both qubit lines. This implies that the $P$-gate commutes with global spin flips $X\otimes X$. Another useful identity is the commutation of two $P$-gates acting on three qubits.

The previous gate identities gave us a better understanding of its action and properties, but the $P$-gate is still represented as a ``black-box'' in this networks. Hence a quantum network model in terms of known gates will be very useful. An equivalent network for the $P$-gate is given in Fig.~\ref{gate_id}(f). An ancilla initialized in the $\ket{0}$ state is coupled to both qubits via two CNOTs and then measured. For the basis states $\ket{xy},\ x,y=0,1$, the measured value of the ancilla corresponds to the parity of the two qubits, $p=x\oplus y$. We stress that this model is not intended as a way of constructing the parity gate out of CNOT gates (since we use the $P$-gate as a primitive resource to replace the CNOT), but only to have a better insight of its action and properties. Using this network model the previous gate identities can be immediately derived.

\section{The parity gate as an entangler}

\subsection{Bell states}

We now turn to the problem of generating entanglement with the $P$-gate. First we define the Bell states (in the following we will consistently omit normalization factors)
\be
\ket{B_{ij}}:= \ket{0}\ket{i}+ (-1)^j \ket{1}\ket{i+1}
\label{bell}
\ee
with $i,j=0,1$. Hence $\ket{B_{00}}= \ket{00}+\ket{11}=\Phi^+$, $\ket{B_{01}}= \ket{00}-\ket{11}=\Phi^-$, $\ket{B_{10}}= \ket{01}+\ket{10}=\Psi^+$ and $\ket{B_{11}}= \ket{01}-\ket{10}=\Psi^-$; $i$ is the {\em parity} bit and $j$ the {\em sign} bit.

One way of entangling two qubits is to use the quantum network for the $CZ$ gate (Fig.~\ref{cz_gate}, top) by adding three Hadamard gates $H$ (two on the target and one on the control qubit). In terms of resources, this procedure is expensive, as it uses one ancilla, six $H$ gates and two $P$-gates, plus measurement of the ancilla and post-processing.

However, if our goal is to entangle two qubits which are in a known separable state, e.g, $\ket{00}$, there exists a simpler way which requires only two Hadamards followed by a parity measurement (and no ancilla). By applying $H\otimes H$ to $\ket{00}$ we obtain an equal superposition of all basis states, which subsequently are projected, using the $P$-gate, on one of the Bell states $\Phi^+, \Psi^+$. More formally, the transformations are:
\ba
\nonumber
\ket{00}\ra H^{\otimes 2} \ket{00}&=& (\ket{00}+\ket{11})_{p=0}+ (\ket{01}+\ket{10})_{p=1} \\
&\stackrel{P}{\ra}& \ket{0}\ket{p}+\ket{1}\ket{p+1}= \ket{B_{p0}}
\ea
where $p=0,1$ is the result of the parity measurement on the state $H^{\otimes 2} \ket{00}$. Notice that although in this case we (randomly) obtain one of the two maximally entangled states with equal probability, we know exactly which one (due to the parity bit $p$), hence we can always end up with a chosen state (say $\Phi^+$) by applying a local post-processing gate $\bbbone\otimes X^p$ (since $\ket{p}= X^p \ket{0}$); here post-processing plays a similar role as in teleportation (see below). It is straightforward to see that all four Bell states $\{ \ket{B_{ij}} \}$ can be {\em deterministically} produced in this way, by starting with one of the basis states $\ket{xy},\ x,y=0,1$ and applying the above procedure, plus post-processing. Since the action of the Hadamard is $H\ket{x}= \ket{0}+ (-1)^x \ket{1},\ x=0,1$, we obtain the following transformation for a basis state $\ket{xy}$
\ba
\nonumber
\ket{xy}\ra H^{\otimes 2}\ket{xy} &\stackrel{P}{\ra}& (-1)^{py} [\ket{0}\ket{p}+ (-1)^{x+y}\ket{1}\ket{p+1}] \\
&=& (-1)^{py} \ket{B_{p, x+y}}
\ea
where $p$ is the value of the measured parity.

\begin{figure}
\putfig{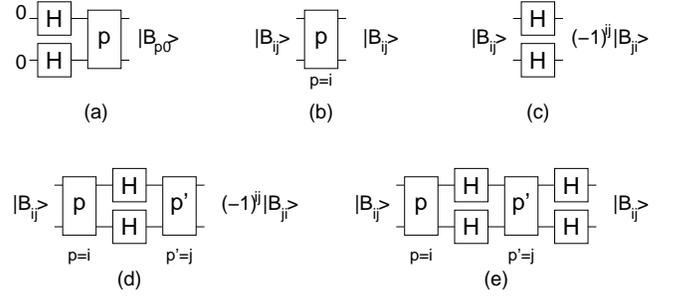}{8.5}
\caption{(a) Producing Bell states with the parity gate. After the $P$-gate, the output state is $\ket{B_{p0}}= \ket{0}\ket{p}+ \ket{1}\ket{p+1}$, where $p$ is the result of the parity measurement. (b) The $P$-gate leaves invariant the Bell states and outputs $p=i$. (c) $H^{\otimes 2}$ maps Bell states into Bell states. (d) A quantum network for a Bell state analyzer; the first $P$-gate measures $i$, then $H^{\otimes 2}$ swaps $i$ and $j$ and the final $P$-gate measures $j$. (e) Adding to the previous network $H^{\otimes 2}$ leaves the input Bell state $\ket{B_{ij}}$ invariant.}
\label{bell_states}
\end{figure}

\subsection{Bell state analyzer and teleportation}

We have seen in the last section that a better strategy to create entangled states is to have ``native'' quantum networks based directly on $P$-gates, instead of translating the standard network by constructing first the CNOT gates out of $P$-gates. Here we develop further this idea by analysing another important quantum protocol, namely teleportation. But first we have to discuss an essential ingredient.

In order to construct a Bell state analyzer, two observations will help. First, it is easy to see that the $P$-gate leaves invariant a Bell state $\ket{B_{ij}}$ and outputs its parity bit $p=i$. Second, the action of $H^{\otimes 2}$ on a Bell state is:
\be
H^{\otimes 2} \ket{B_{ij}}= (-1)^{ij} \ket{B_{ji}}
\ee
hence $H^{\otimes 2}$ maps Bell states into Bell states and swaps $i$ with $j$ (up to a phase). From this two observations if follows immediately that a {\em non-destructive} Bell state analyser can be constructed with two $P$-gates and two Hadamards as in Fig.~\ref{bell_states}(d). The first $P$-gate measures $i$, then $H^{\otimes 2}$ swaps $i\leftrightarrow j$ and the final $P$-gate measures $j$. This network leaves two of the Bell states invariant and swaps the other two. If we want to leave invariant all four Bell states, we have to add another two Hadamards as in Fig.~\ref{bell_states}(e).

The same resource counting argument as before can be applied here. A projective Bell measurement involves a CNOT gate followed by a Hadamard on the control qubit and a measurement of both qubits. The ``native'' scheme proposed above is indeed simpler than one based on implementing first the CNOT gate: it involves two $P$-gates and two (instead of six) Hadamards; moreover, no ancilla and no final measurement is needed.

It is worth mentioning that the Bell state analyzer discussed here differs from the one proposed in \cite{spin_parity} in several respects. First, the present design is nondestructive, i.e., the Bell-states are left invariant after measurement. Second, the network in Fig.~\ref{bell_states} works for generic qubits, in contrast to \cite{spin_parity} which is designed for spin qubits. Finally, in our case we use only two $P$-gates (instead of three).

\begin{figure}
\putfig{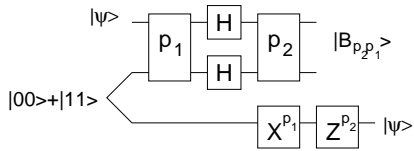}{5.5}
\caption{A teleportation circuit using two $P$-gates and two Hadamards; note that no ancilla and no final measurement of the Alice's two qubits is needed. After the measurement, Alice remains with a Bell state $(-1)^{p_1 p_2} \ket{B_{p_2p_1}}$. Bob recovers $\ket{\psi}$ by applying a local transformation $Z^{p_2} X^{p_1}$.}
\label{teleport}
\end{figure}

Let us now turn to teleportation. The protocol is well-known; first, Alice and Bob share an EPR state. In order to teleport to Bob the unknown state $\ket{\psi}= a\ket{0}+ b\ket{1}$, Alice performs a projective Bell measurement on $\ket{\psi}$ and her half of the EPR pair, then sends to Bob two bits of classical information $(i,j)$. Bob then applies to his state one of the four unitaries $U_{ij}$. Using the Bell state measurement discussed previously, we obtain the network shown in Fig.~\ref{teleport}. A straightforward calculation shows that the action of this network is:
\ba
\nonumber
&&(a\ket{0}+ b\ket{1})_A (\ket{00}+\ket{11})_{AB} \\
&\stackrel{P_2 H^{\otimes 2} P_1}{\longrightarrow}& \ket{\phi}_A \otimes X^{p_1} Z^{p_2}(a\ket{0}+ b\ket{1})_B
\ea
where after the teleportation Alice ends up with a Bell state $\ket{\phi}_A= (-1)^{p_1p_2}\ket{B_{p_2 p_1}}$ and Bob's qubit becomes $X^{p_1} Z^{p_2}(a\ket{0}+ b\ket{1})_B$. Therefore, after Alice sends him the results of her parity measurements $(p_1,p_2)$, Bob recovers the unknown state $\ket{\psi}$ by applying a local unitary $Z^{p_2} X^{p_1}$.

It is important to note the double role played here by the parity measurements. On one hand, they entangle the unknown qubit $\ket{\psi}$ with one of the EPR pair. On the other, the two $P$-gates also provide the two bits of classical information Alice sends to Bob in order to recover the teleported state. As a result, Alice does not need to perform the final measurement on her two qubits. Moreover, after the protocol Alice still has an entangled state.

\subsection{Multi-qubit entanglement}\label{n_qubits}

We can generalize the scheme for producing Bell states (Fig.~\ref{bell_states}(a)) to an arbitrary number of qubits. Starting with a separable $n$-qubit state $\ket{0}^{\otimes n}$, we put it in an equal superposition of all basis states by applying Hadamards to each qubit, $\ket{+}^{\otimes n}= H^{\otimes n}\ket{0}^{\otimes n}= \sum_{i=0}^{2^n-1} \ket{i}$. By performing $n-1$ parity measurements ($P$-gates) between next-neighbour qubits, we obtain a state locally equivalent to a $n$-GHZ state $\ket{GHZ_n}= \ket{0}^{\otimes n}+ \ket{1}^{\otimes n}$. Let us see this in more detail.

Assume the parity measurements are given by the vector ${\bf p}=(0, p_2,..,p_n)$, where $p_i$ is the value of the parity for the $(i-1,i)$ pair of qubits. Construct the binary vector ${\bf j}=(0, j_2,..,j_n)\in \bbbz_2^n$, with $j_i= j_{i-1}\oplus p_i$. After the $n-1$ parity measurements on the state $\ket{+}^{\otimes n}$, the only surviving terms in the equal superposition sum are $\ket{j}:=X({\bf j}) \ket{0}^{\otimes n}$ and $X^{\otimes n}\ket{j}$, since these are the only terms compatible with the outcome of the $n-1$ parity measurements given by $\bf p$. Therefore the output state after the parity measurements is
\be
\ket{\psi_p}= X({\bf j}) (\bbbone+X^{\otimes n}) \ket{0}^{\otimes n}= X({\bf j}) \ket{GHZ_n}
\ee
Thus by applying $n-1$ (since $j_1=0$ always) post-processing spin-flips $X({\bf j})$, we obtain $\ket{GHZ_n}$ as the final state.

\begin{table}[t]
\caption{Resource counting for a $n$-GHZ state in two models, the ``native'' one (employing directly $P$-gates) and the CNOT-based implementation.}
\begin{ruledtabular}
\begin{tabular}{l|cr}
 & ``native'' & CNOT-based \\
\hline
ancill\ae & 0 & $n-1$ \\
measurements (ancill\ae) & $0$ & $n-1$ \\
$P$-gates & $n-1$ & $2(n-1)$ \\
Hadamards & $n$ & $5n-4$ \\
post-processing & $n-1$ & $2(n-1)$ \\
\end{tabular}
\end{ruledtabular}
\label{tab1}
\end{table}

It is instructive to compare the resources for producing a $n$-GHZ state using the above method against the traditional method (1 Hadamard gate on the first/control qubit plus $n-1$ CNOT gates between the first and the remaining qubits). The results are presented in Table \ref{tab1}.

\subsection{Cluster state fusion with $P$-gates}

There has been recently a growing interest in cluster state model in the context of linear optics QC \cite{nielsen2,browne,benjamin,benjamin2,kieling,bk,beige}. The main idea behind these proposals is to grow (or fuse) cluster states using only linear optics methods, hence bringing together the KLM proposal and the cluster state QC. Since the linear optics methods are necessarily probabilistic, several strategies for fusing smaller cluster states into larger one have been proposed in order to optimize the output \cite{browne,kieling}.

The scheme for generating $n$-GHZ states presented above has an interesting interpretation as a fusion operation of graph states (a GHZ state is locally equivalent to a star graph state). Suppose we have two GHZ states, $\ket{GHZ_n}$ and $\ket{GHZ_m}$, with $m \le n$. Performing a single parity measurement between a qubit of the first state and one of the second, we obtain a state locally equivalent to a $\ket{GHZ_{n+m}}$ state:
\be
\ket{GHZ_n}\otimes \ket{GHZ_m} \stackrel{P}{\rightarrow} U^p \ket{GHZ_{n+m}}
\ee
where $p=0,1$ is the result of the parity measurement and $U:=\bbbone_n \otimes X^{\otimes m}$. The parity fusion is deterministic, i.e., we always obtain a state locally equivalent to $\ket{GHZ_{n+m}}$. This is to be expected, since we know we can construct a deterministic $CZ$ gate out of $P$-gates \cite{spin_parity}.

In order to put this result in perspective, it is worth giving a brief overview of cluster state QC. Raussendorf and Briegel \cite{1wqc} introduced the one-way quantum computing model (i.e., cluster state QC) as an alternative to the standard quantum network paradigm, to which is polynomially equivalent. They showed that one can realize any quantum algorithm by performing only single-qubit measurements on a highly entangled initial state, the {\em cluster state}. A cluster state is thus a universal resource for quantum computing and corresponds to a graph state associated to a 2D square lattice.

Cluster state QC has a distinctive feature compared to the classical quantum network model, namely all the entanglement necessary in the computation is already present in the initial cluster state. Once this state is constructed, the desired algorithm is performed by a series of single qubit measurements (and possible feed-forward single-qubit gates). Also, an advantage is that we can use non-deterministic gates and feed-forward in order to construct the initial cluster state.

Let $G=(V,E)$ be a (simple) graph; $V$ and $E$ are, respectively, the set of vertices and edges. The graph state $\ket{G}$ is defined as follows. To each vertex $v\in V$ we associate a qubit in the initial state $\ket{+}= \ket{0}+\ket{1}$. For each edge $(i,j)\in E$ we apply a $CZ_{ij}$ operation between the corresponding qubits $i,j$. The associated graph state $\ket{G}$ is then:
\be
\ket{G}= \prod_{(i,j) \in E} CZ_{ij} \ket{+}^{\otimes n}
\ee
Any stabilizer state is locally equivalent to a graph state (and vice-versa). Since a square lattice (in any dimension) corresponds to a bipartite (i.e., two-colorable) graph, it follows that any cluster state is also a two-colorable graph state, hence it is also locally equivalent to a Calderbank-Shor-Steane (CSS) state \cite{hiz3,lo}. Several known entangled states are locally equivalent to graph states; a notable example is the $n$-GHZ state, which corresponds to an $n$-vertex star graph.

As we have seen before, parity measurements enable fermionic linear optics QC, and moreover this can be performed deterministically.

One way to construct a cluster state with parity gates is to apply a $CZ$ gate for each edge $(i,j)\in E$ via the construction Fig.~\ref{cz_gate}, as in \cite{feng}. However, each $CZ$ gate requires extra resources: one ancilla, two $P$-gates, Hadamards, ancilla measurement and post-processing (see Fig.\ref{cz_gate}). Can we do better than this? As before, one would expect that a ``native'' network would avoid this resource overhead.

Assume we have a cluster state $\ket{G_0}$ and we apply a $P$-gate on two of its qubits which are initially disjoint, i.e., $(1,2)\not\in E$. Without loss of generality, the initial cluster state can be made of two disjoint clusters and the operation will then correspond to a cluster state fusion. For simplicity, we label the two qubits on which we apply the $P$-gate as 1 and 2. Following Benjamin \cite{benjamin}, we can write the initial cluster state as:
\be
\ket{G_0}= \left( \ket{0}+ \ket{1}Z^{(1)} \right) \left( \ket{0}+ \ket{1}Z^{(2)} \right) \ket{G'}
\ee
where $Z^{(a)}= \prod_{(a, k) \in E} Z_k$, $a=1,2$ are the product of $Z$ operators applied to all qubits having a common edge with qubit $a$; the state $\ket{G'}$ denotes the state of all other qubits. Applying a parity projective measurement to the first two qubits ($p$ is the measured parity), we obtain (since we assumed $(1,2)\not\in E$)
\be
\ket{G_1}= X_2^p [Z^{(2)}]^p \left( \ket{00}+ \ket{11}Z^{(1)} Z^{(2)} \right) \ket{G'}
\ee
We now apply a Hadamard on the first qubit. In the final cluster state, qubit 2 inherits all the edges (modulo 2) of qubit 1, which now becomes a ``leaf'', i.e., linked only to qubit 2:
\be
\ket{G_2}= X_2^p [Z^{(2)}]^p (\ket{0}+ \ket{1}Z_2) (\ket{0}+ \ket{1} Z^{(1)} Z^{(2)}) \ket{G'}
\ee
Hence a $P$-gate produces a fusion of two cluster states as in \cite{benjamin}, with the added advantage that it is always successful, since both outcomes of the parity measurement produce locally equivalent states.

\section{The parity gate in hybrid quantum networks}

\subsection{General setup}

So far we have discussed the $P$-gate in the context of generic qubits. The original proposal of Beenakker {\em et al.} \cite{spin_parity} was designed as a spin CNOT gate using charge-to-spin conversion and charge parity measurements (Fig.~\ref{cz_gate}). The reason comes from the fact that spin-spin interaction is much weaker than charge-charge (Coulomb) interaction, hence it is more difficult to perform directly a spin $P$-gate. On the other hand, spin degrees of freedom make good qubits, as they have a much longer decoherence time than charge (mode) degrees of freedom. This motivates us to investigate {\em hybrid quantum networks}, i.e., quantum circuits containing both spin and mode degrees of freedom.

In this section we build on the interplay between charge and spin qubits and we explore various schemes for entanglement generation using the charge $P$-gate. The main result is a novel method of entangling spins using only charge parity measurements and linear gates (beam splitters and spin flippers), without using any spin-spin interaction.

We assume we have an array of $n$ quantum particles, each with two degrees of freedom; we will refer to the internal degree of freedom as ``spin'', and to the external one as ``mode''. For definiteness we will call such a particle ``electron'', although the architecture described here is completely general and encompasses several possible real-life implementations. Some examples are static electrons in double quantum dots, mobile electrons in parallel quantum wires or atoms in optical lattices (in a Mott insulator phase). The only requirements are: (i) the single-particle Hilbert space factorizes in spin and mode degrees of freedom and (ii) apart from single-qubit gates (on both spin and mode qubits), we are able to implement $P$-gates on the mode qubits (hereafter referred as the ``charge parity gate'') between two neighbouring particles.

In this setup each electron encodes two qubits: a spin qubit, defined by the spin state ($\up$ or $\down$), and a mode (charge) qubit, defined by the spatial mode (0 or 1). Thus the Hilbert space of a single electron is ${\cal H}= {\rm span} \{ \ket{\up 0}, \ket{\up 1}, \ket{\down 0}, \ket{\down 1} \}$. For clarity we will use sometimes the subscripts $\sigma\, (k)$ to indicate the spin (mode) qubits.

Several authors have investigated projective measurements either as a way to achieve universality \cite{virmani} or as a means of entangling electrons in mesoscopic devices \cite{mao,ruskov,ruskov2, stace1,stace2,feng}. An extensive study of projective measurements of electrons in double quantum dots has been made in \cite{stace1, stace2}, in which various aspects of the geometry involved and the effect of asymmetry were discussed.

\subsection{A cheap resource: single-particle entanglement}

As discussed previously, each electron encodes two qubits and therefore we can have single-particle entanglement between spin and charge degrees of freedom of the same electron. This is a ``cheap'' resource, as it can be produced with linear elements: a beam-splitter and a local spin-flip. To see this, assume we start with the electron in the basis state $\ket{\up 0}_{\sigma k}$. By applying a Hadamard on the mode qubit followed by a spin flip {\em exclusively} on mode 1 \footnote{More generally, an arbitrary spin gate $U_\sigma$ applied only to mode 1 is equivalent to a controlled-$U(k,\sigma)$ gate between mode (control) and spin (target) qubits.}, we obtain:
\be
\ket{\up 0}_{\sigma k} \ra \ket{\up}(\ket{0}+\ket{1}) \ra \ket{\up 0}+ \ket{\down 1}= \ket{EPR}_{\sigma k}
\ee
i.e., a spin-mode entangled EPR state of a single electron.

A natural question is how can we use this easy-to-prepare, single-particle entanglement as a resource to entangle different particles. For example, of particular interest for implementations would be to entangle only the spin degrees of freedom of several particles.

\begin{figure}
$\insertfig{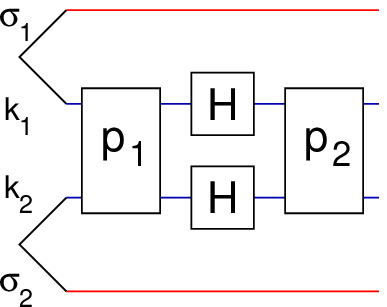}{2.5}$ \hspace{1.2cm} $\insertfig{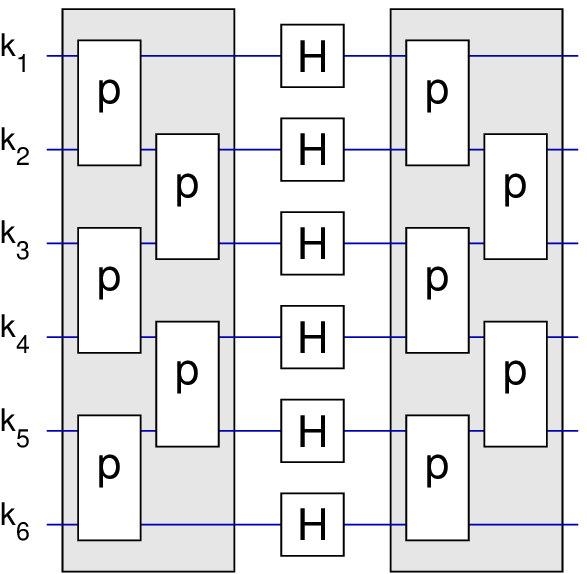}{3.5}$
\caption{(color online) Left: Entanglement swapping between spin (red) and charge (blue) qubits. The initial state contains only single-particle entanglement (both electrons are entangled in spin-charge) $(\ket{\up 0}+ \ket{\down 1})^{\otimes 2}$. By performing a Bell state measurement on the charge qubits we obtain a hyper-entangled state of two electrons (entangled in both spin and charge) $\ket{EPR}_{\sigma} \ket{EPR}_k$. Right: Generalizing to $n$ qubits; for simplicity, only the mode qubits are depicted, as no operation on spins is performed. The initial state is $(\ket{\up 0}+ \ket{\down 1})^{\otimes n}$; the output of the quantum network is $\ket{GHZ_n}_{\sigma} \ket{GHZ_n}_k$.}
\label{e_swap}
\end{figure}

Let us first analyse the simple case of two electrons. The quantum network representing the four qubits encoded by the two particles is given in Fig.~\ref{e_swap}, left; the top (bottom) two qubit lines represent the degrees of freedom ($\sigma$ and $k$) of the first (second) electron, and both electrons are spin-mode entangled. Now it is easy to see the solution of the problem: perform an entanglement swapping operation (using a Bell state measurement) on the mode qubits of the two particles. As a result, the final state will be a hyper-entangled state of two electrons (entangled in both spin and charge). The gate sequence is the following:
\be
(\ket{\up 0}+ \ket{\down 1})^{\otimes 2} \stackrel{P}{\ra} \ket{GHZ_4}_{\sigma k} \stackrel{P H^{\otimes 2}}{\longrightarrow} \ket{EPR}_{\sigma} \ket{EPR}_k
\label{epr}
\ee

It is worth noting the intermediate state in the above equation. After applying the first $P$-gate to the initial state, we obtain a 4-GHZ state (e.g., $\ket{\up \up 00}+ \ket{\down \down 11}$) containing genuine 4-qubit entanglement. This is different from the entanglement of the final state $\ket{EPR}_{\sigma} \ket{EPR}_k \simeq_{LU} (\ket{\up \up}+ \ket{\down \down})_\sigma (\ket{00}+ \ket{11})_k$. However, from the point of view of implementations, the second state is preferable, as it does not mix spin and charge degrees of freedom. One can use the final state in eq.~(\ref{epr}) in a spin-qubit processor by measuring the mode qubits and leaving thus only spin-spin entanglement between the two electrons.

\subsection{Entangling more spins}

We now show how to generalize the previous construction to an arbitrary number of particles. Assume we start with $n$ electrons in the initial state $(\ket{\up 0}+ \ket{\down 1})^{\otimes n}$. As before, this state contains only single-particle entanglement (between spin and mode degrees of freedom) and can be produced with linear elements (beam-splitters and local spin-flips). The network generalizing the entanglement swapping is shown in Fig.~\ref{e_swap}, right. Denote by $P_{n-1}$ the set of $n-1$ charge-parity measurements (grey boxes in the figure). Then the transformation performed by the network is:
\ba
\nonumber
(\ket{\up 0}+ \ket{\down 1})^{\otimes n} &\stackrel{P_{n-1}}{\longrightarrow}_{LU}& \ket{GHZ_{2n}}_{\sigma k} \\
&\stackrel{P_{n-1} H^{\otimes n}}{\longrightarrow}_{LU}& \ket{GHZ_n}_{\sigma} \ket{GHZ_n}_k
\label{ghz_n}
\ea
The first $P_{n-1}$ gate projects the initial state to the $n$-particle hyper-entangled state $\ket{GHZ_{2n}}_{\sigma k}$ (up to local unitaries). This state contains entanglement between spin and mode degrees of freedom of all electrons; as such, it is appealing from a conceptual point of view, e.g., to test Bell inequalities for many particles and between different degrees of freedom, $\sigma$ and $k$. 

However, the state is not practical as a computational resource, since the decoherence time of spin and mode degrees of freedom are different by a few order of magnitudes. By further applying $P_{n-1} H^{\otimes n}$ we disentangle the spin and mode degrees of freedom and we obtain a state locally equivalent to a $n$-GHZ spin state, which can be used as a computational resource. Although the exact state depends on the outcomes of the parity measurements, the output always contains genuine $n$-particle entanglement and can be transformed {\em deterministically} into the standard state $\ket{GHZ_n}:= \ket{0}^{\otimes n}+ \ket{1}^{\otimes n}$ by local transformations only.

With the notation $\ket{\up}= \ket{0}$, $\ket{\down}= \ket{1}$, the initial state can be written as:
\be
\ket{\psi_0}= (\ket{\up 0}+ \ket{\down 1})^{\otimes n}= \sum_{i=0}^{2^n-1} \ket{i}_\sigma \ket{i}_k
\ee
and in the last sum we have clustered separately the spin and mode qubits. We now evaluate the transformations step-by-step.

1. Assume the result of the first $P_{n-1}$ gate is given by the (charge) parity vector ${\bf p}:=(0, p_2, ..., p_n)$; as before, $p_i$ is the result of the parity of the $(i-1,i)$-pair of qubits. Let ${\bf j}:=(0, j_2, ..., j_n)$, with $j_i=j_{i-1}\oplus p_i$. For the mode qubits, the only basis states compatible with the outcome of the measurements are $\ket {j}_k$ and $X^{\otimes n} \ket {j}_k$. Thus, the state becomes:
\ba
\nonumber
\ket{\psi_1}&=& (\bbbone + X^{\otimes 2n}) \ket{j}_\sigma \ket {j}_k= X({\bf j})_\sigma X({\bf j})_k  (\Uparrow {\bf 0} + \Downarrow {\bf 1}) \\
&=& X({\bf j})_\sigma X({\bf j})_k \ket{GHZ_{2n}}
\ea
where $\Uparrow:= \ket{\up}^{\otimes n}$, ${\bf 0}:= \ket{0}^{\otimes n}$ etc. 

2. Apply a global $H_k^{\otimes n}$ on the mode qubits. Since $HX=ZH$ and $H\ket{x}= \ket{0}+ (-1)^x \ket{1}$ we obtain
\ba
\nonumber
\ket{\psi_2}&=& X({\bf j})_\sigma Z({\bf j})_k  \left[ \Uparrow (\ket{0}+\ket{1})_k^{\otimes n} + \Downarrow (\ket{0}- \ket{1})_k^{\otimes n} \right] \\
\nonumber
&=& X({\bf j})_\sigma Z({\bf j})_k \left[ (\Uparrow+ \Downarrow)\sum_{i\, even} \ket{i}_k+ (\Uparrow- \Downarrow) \sum_{i\, odd} \ket{i}_k \right] \\
\label{psi2}
\ea
In the last equation the two sums are over even (respectively, odd) terms $i$. The parity of a binary vector is defined as $\pi({\bf v})= \sum_i v_i \mod 2$; thus ${\bf v}$ is even (odd) if it has an even (odd) number of '1' entries.

This definition can also be applied to basis states $\ket{j}$. Since $X^{\otimes n}$ flips all the '0' and '1' in the binary representation of $j$, we have $\pi(\ket{j})+ \pi(X^{\otimes n} \ket{j})= n \mod 2$. Therefore the basis vectors $\ket{j}$ and $X^{\otimes n} \ket{j}$ have the same (opposite) parity if the number of qubits $n$ is even (odd).

3. Now apply the second $P_{n-1}$ gate. Since the $P$ gate commutes with $Z({\bf j})_k$, we can apply it directly to the last bracket in eq.~(\ref{psi2}). Let ${\bf p'}=(0,p'_2,..,p'_n)$ be the new parity vector and ${\bf m}=(0, m_2,..,m_n)$, with $m_i= m_{i-1}\oplus p'_i$. After the parity measurement on the mode qubits, the only surviving terms in the last bracket are $\ket{m}_k$ and $X^{\otimes n} \ket{m}_k$. To simplify the calculation, we assume that the number of qubits $n$ is even \footnote{If $n$ is odd, we need to apply an extra $CZ_1(\sigma, k)$ gate on the the first electron. As discussed above, this is equivalent to a local spin $Z_1$ gate on mode-1 only. Thus the assumption $n$ even simplifies the procedure.}. In this case $\ket{m}_k$ and $X^{\otimes n} \ket{m}_k$ have the same parity and the final state is:
\ba
\nonumber
\ket{\psi_3}&=& X({\bf j})_\sigma Z({\bf j})_k \left( \Uparrow + (-1)^{\pi(m)}\Downarrow \right)_\sigma \left(\ket{m}+ X^{\otimes n}\ket{m} \right) _k \\
&=& X({\bf j})_\sigma Z_{1,\sigma}^{\pi(m)} Z({\bf j})_k X({\bf m})_k \ket{GHZ_n}_\sigma \ket{GHZ_n}_k
\ea

Let us discuss some properties of our scheme. First, it is {\em deterministic}, as the final state can be locally transformed, with 100\% success, in $\ket{GHZ_n}_\sigma \ket{GHZ_n}_k$, irrespective of the result of the parity measurement. Second, it is highly parallelisable. All the steps discussed above (initial state preparation, $H^{\otimes n}$ and $P_{n-1}$ gates) can be done in parallel. Moreover, apart from the preparation of $\ket{\psi_0}$ (which requires $n$ local spin flips), all other operations (Hadamards and $P$-gates) act only on mode qubits.

The above spin $n$-GHZ state can be used as a resource in a spin quantum computer or as a precursor for constructing more general cluster states.

\subsection{A new spin-$CZ$ gate}

As we have seen, the ``native'' quantum networks investigated so far simplify the resources required for entanglement generation, teleportation and Bell state analysis. However, they have a drawback -- they are designed for a specific purpose and, in the case of the various entanglement generation schemes, they start from a given (separable) initial state.

On the other hand, the spin-$CZ$ gate in Fig.~\ref{cz_gate} acts coherently on arbitrary input states, i.e., it preserves the general superposition of the input \cite{spin_parity}. As such, it can be used in general quantum algorithms. The key element in this construction is the use of an ancilla prepared in the $\ket{0}+ \ket{1}$ state, acting as an ``encoder''. Parity measurement is done sequentially on control-ancilla and ancilla-target qubits (with a $H$ gate in-between) and this strategy preserves the coherence of the input qubits. However, since a spin-parity gate is not directly feasible, one uses a charge-parity gate and a charge-to-spin conversion via two polarizing beam-splitters (PBS), adding thus an extra layer of complication.

Counting the resources for the spin $CZ$ gate in Fig.~\ref{cz_gate}, we obtain: two charge-parity gates, four PBS (two for each spin-parity gate), three spin-Hadamards, a spin-ancilla and a spin measurement (plus post-processing).

The existence of the spin ancilla implies an extra particle (e.g., electron in a quantum dot/wire). This means that for an $n$-qubit array we need an extra $n-1$ ancill\ae\ placed between the computational qubits, effectively doubling the resources.

A natural question arises: Is there a way to simplify the design of the $CZ$ gate, without compromising its coherent action on arbitrary superpositions?

\begin{figure}
$\insertfig{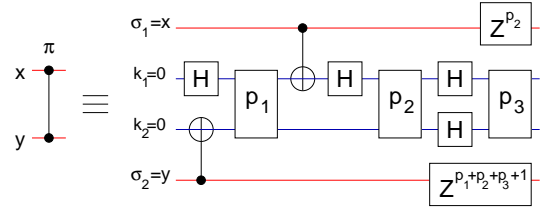}{7}$
\caption{(color online) The new spin-$CZ$ gate using the mode qubits as ancill\ae.}
\label{cz_new}
\end{figure}

From the above discussion two ideas should be stressed: the need of an ancilla (in order to preserve the superpositions) and the use of charge-parity detectors (experimentally feasible). These two points help us to re-frame the above question in a different way: Can we employ the mode degrees of freedom as ancill\ae, since they are already used for the charge-parity measurements? The answer is yes, and the new spin $CZ$ gate is shown in Fig.~\ref{cz_new}.

It involves only two electrons, with spin degrees of freedom encoding the computational qubits and mode degrees of freedom the ancill\ae; the mode qubits are initialized to the $\ket{00}$ state. The new scheme has some desirable features:

- it eliminates completely the spin ancilla (hence the extra electron);

- it replaces a spin measurement (ancilla) with a charge-parity gate; this is an advantage, as spin measurement is notoriously difficult to implement in practice and usually requires spin-to-charge conversion;

- it uses only two (instead of four) PBS (the CNOT gates between $\sigma$ and $k$);

- it has four (instead of three) $H$ gates, but they are on the mode (instead of spin) degree of freedom, which make them easier to perform. A Hadamard on mode is equivalent to a beam-splitter, which is easier from a practical point of view than a $H$ on spin (it requires local magnetic fields or Rashba-active regions).

The action of the network in Fig.~\ref{cz_new} on the initial state $\ket{x00y}:= \ket{x}_{\sigma_1}\ket{0}_{k_1} \ket{0}_{k_2}\ket{y}_{\sigma_2}$ is:
\be
\ket{x00y} \ra (-1)^{xy} (-1)^{p_1 p_2} \ket{x}_{\sigma_1} \ket{B_{p_3 p_2}}_{k_1 k_2} \ket{y}_{\sigma_2}
\ee
The $(-1)^{p_1 p_2}$ factor is independent of the input state $\ket{xy}$ and can be neglected (in complete analogy to the original $CZ$ gate, see \cite{spin_parity}). Thus the above gate performs a $CZ$ gate on spins, as desired. It can be shown that the new $CZ$ gate acts coherently, hence it preserves arbitrary superpositions of the input spin state. Also, due to the post-processing step (the two $Z^p$ gates on spin), the gate is {\em deterministic}, hence it works with 100\% probability. 

It is important to note that the mode qubits end up in an entangled Bell state. If required, this can be corrected by a projective measurement on one of the $k$-qubits and a post-processing stage, in order to bring it to $\ket{00}_{k_1 k_2}$ state.

\section{Conclusions}

Recently there has been an increased interest in parity measurements as a resource for fermionic quantum computation with linear elements. This idea is attractive since it reduces the stringent requirements needed to perform two-qubit gates (like accurate control of spin-spin interaction and gate timing) to a conceptually easier problem: a projective charge-parity measurement.

One of the motivations of this article was to develop a toolbox for the $P$-gate in the context of quantum networks. This framing of the problem enabled us to derive gate identities for $P$-gates and novel methods for generating entanglement in $n$ particle systems. We have shown that designing native quantum networks for $P$-gates is more efficient, in terms of resources, than building CNOT gates out of $P$-gates. Examples include a novel Bell-state analyser and a fast method to construct $n$-GHZ states.

Cluster states are an essential resource in the one-way QC model of Raussendorf and Briegel \cite{1wqc}. One way of constructing these states are by performing a fusion operation between two graph states in order to obtain a larger one. We have shown that $P$-gates can perform such fusion operations deterministically.

We then extended our analysis to hybrid quantum networks, containing both spin and mode qubits. In this case we showed that one can entangle spins by performing charge parity measurements on a state containing only single-particle entanglement. Finally, we constructed a new spin $CZ$ gate by using charge $P$-gates and mode qubits as ancill\ae.


\end{document}